\documentclass[onecolumn]{revtex4}

\usepackage{graphicx}
\usepackage{dcolumn}
\usepackage{color}
\usepackage{bm}

\input epsf

\begin{document}

\title{\Large Bulk scalar field in warped extra dimensional models}

\author{\bf Sumanta
Chakraborty \footnote{sumantac.physics@gmail.com} }

\affiliation{IUCAA, Post Bag 4, Ganeshkhind,
Pune University Campus, Pune 411 007, India}

\author{\bf Soumitra SenGupta \footnote{tpssg@iacs.res.in}}

\affiliation{Department of Theoretical Physics, Indian Association for the Cultivation of Science,
Kolkata-700032, India}

\date{\today}

\begin{abstract}
This work presents a general formalism to analyze a generic bulk
scalar field in a multiple warped extra-dimensional model with
arbitrary number of extra dimensions. The Kaluza-Klein mass modes
along with the self-interaction couplings are determined and the
possibility of having lowest lying  KK mode masses near TeV scale
are discussed. Also some numerical values for low-lying KK modes has been
presented showing explicit localization around TeV scale.
It is argued that the appearance of large number of closely
spaced KK modes with enhanced coupling
may prompt possible new signatures in collider physics.
\end{abstract}

\maketitle

\section{Introduction}\label{bmintro}
Theories with extra spacetime dimensions have drawn considerable
attention ever since the original proposal by Kaluza and Klein.
There has been renewed interest in such theories since the
emergence of string theory. Several new ideas in this context have
been proposed and have interesting consequences for particle
phenomenology and cosmology
\cite{Hamed1,Antoniadis,Horava,Lukas,Hamed2}. In these higher-dimensional 
models, spacetime is usually taken to be a product of a
four-dimensional spacetime and a compact manifold of dimension
$n$. While gravity can propagate freely through the extra
dimensions, Standard Model particles are confined to the four
dimensional spacetime. Observers in this three spatial-dimensional wall (a
``3-brane") will measure an effective Planck scale
$M_{pl}^{2}=M^{n+2}V_{n}$, where $V_{n}$ is the volume of the
compact space. If $V_{n}$ is large enough it could make Planck
scale of the order of TeV, thus removing the hierarchy between the
Planck and the weak scale.

Subsequently, Randall \textit{et al.} \cite{Randall1,Randall2}
proposed a higher-dimensional scenario that is based on
nonfactorizable geometry and accounts for the hierarchy
without introducing large extra dimensions. However, the
braneworld model itself is not stable and it was shown in
Ref. \cite{Wise1} that by introducing a
scalar field in the bulk, the modulus-namely the brane
separation in the RS model-can be stabilized without any fine-tuning.
Assumption of negligibly small scalar backreaction on
the metric in the GW approach prompted further work in
this direction, where
the modification of the RS metric due to backreaction
of the bulk fields has been derived (see \cite{DeWolfe}).
The stability issues
in such cases have been reexamined for time-dependent
cases \cite{Chakraborty,Csaki}; also the effect of gauge fields
or higher form fields have been studied in several works (see \cite{Das}).

In an effort to search for the signatures of extra dimensions,
the roles of the Kaluza-Klein modes of different bulk fields on the
phenomenology at the standard model brane
are of crucial importance.

For the five-dimensional RS model, \cite{Wise2} determined the
bulk scalar KK modes and their self-interactions. It is found that
due to the exponential redshift factor in the RS model, KK scalar
modes in this spacetime have TeV scale mass splitting and inverse
TeV couplings (see \cite{Randall2}). This is in sharp contrast to
the KK decomposition in product spacetimes, which for large
compactified dimensions, give rise to a large number of light KK
modes (see \cite{Wells}) with a very small coupling with brane
fields. Due to this very distinct feature, the RS model has
interesting consequences \cite{Wise2,Sengupta}.

Motivated by string theory and other extra-dimensional models
where one can have several extra dimensions,
in this paper we extend the results of the bulk scalar
model in five-dimensional RS space-time to arbitrary number of warped dimensions
and have obtained the KK decomposition
of the scalar KK masses. We have shown that in these multiply
warped models we have much larger number of  KK modes than the five-dimensional
RS counterpart with effective couplings in the inverse TeV range.
We have also discussed possible numerical values for various
parameters in our theory and have used them to get
possible numerical values of low-lying KK mode masses
in our multiply warped model, showing explicit localization in TeV brane.
Our results also establish a general formula for determining
these KK masses and couplings in the presence of  any arbitrary number of extra dimensions.

The paper is organized as follows: We give a brief explanation for
six-dimensional doubly warped spacetime in Sec. \ref{bmsix},
the calculation for bulk scalar field has been done in this six-dimensional 
spacetime in Sec. \ref{bmbulk}. The same
calculations have been finally extended  to higher-dimensional
spacetime with arbitrary number of extra dimensions in Secs.
\ref{bmseven} and \ref{bmbulk7}. The paper ends with a short
discussion of our results.
\section{Six-Dimensional Doubly Warped Spacetime and Einstein Equations}\label{bmsix}

In this section we shall discuss doubly compactified six-dimensional 
spacetime with $Z_{2}$ orbifolding along
each compactified direction. For a detailed discussion
we refer the reader to \cite{Sengupta}. The manifold
under consideration is given by,
$M^{1,5}=\left[M^{1,3}\times S^{1}/Z_{2}\right]\times S^{1}/Z_{2}$ \cite{Sengupta}.

We let the compactified dimensions to $y$ and $z$,
respectively. The noncompactified dimensions are
taken to be, $x^{\mu}(\mu =0,1,2,3)$. The moduli
along the compact dimensions are given by $R_{y}$
and $r_{z}$, respectively. The corresponding metric ansatz is taken as
\begin{equation}\label{bm1}
ds^{2}=b^{2}(z)\left[a^{2}(y)\eta _{\mu \nu}dx^{\mu}dx^{\nu}+R_{y}^{2}dy^{2}\right]+r_{z}^{2}dz^{2},
\end{equation}
with $\eta _{\mu \nu}=\textrm{diag}(-1,1,1,1)$. Thus we have four
orbifold fixed points, which are given by
$(y,z)=(0,0), (0,\pi), (\pi ,0), (\pi ,\pi)$, respectively.

The total bulk-brane action could be given by
\begin{eqnarray}\label{bm2}
 && S=S_{6}+S_{5}+S_{4}
\\
&& S_{6}=\int d^{4}xdydz\sqrt{-g_{6}}\left(R_{6}-\Lambda _{6} \right)
\\
&& S_{5}=\int d^{4}xdydz \left[V_{1}\delta(y)+V_{2}\delta(y-\pi)\right] + \int d^{4}xdydz \left[V_{3}\delta(z)+V_{4}\delta(z-\pi)\right]
\\
&& S_{4}=\sum _{i=1}^{2}\sum _{j=1}^{2} \int d^{4}xdydz \sqrt{-g_{vis}}\left(L-V\right)\delta (y-y_{i})\delta (z-z_{j}).
\end{eqnarray}
Here the brane potentials in general have the particular
functional dependence $V_{1,2}=V_{1,2}(z)$ and $V_{3,4}=V_{3,4}(y)$.
Finally the full six-dimensional Einstein's equation is given by,
\begin{eqnarray}\label{bm3}
-M^{4}\sqrt{-g_{6}}\left(R_{MN}-\frac{R}{2}g_{MN}\right)&=&\Lambda _{6}\sqrt{-g_{6}}g_{MN}
\nonumber
\\
&+&\sqrt{-g_{5}}V_{1}(z)g_{\alpha \beta}\delta ^{\alpha}_{M}\delta ^{\beta}_{N}\delta(y)
\nonumber
\\
&+&\sqrt{-g_{5}}V_{2}(z)g_{\alpha \beta}\delta ^{\alpha}_{M}\delta ^{\beta}_{N}\delta(y-\pi)
\nonumber
\\
&+&\sqrt{-\tilde{g_{5}}}V_{3}(y)\tilde{g}_{\tilde{\alpha}\tilde{\beta}}\delta ^{\tilde{\alpha}}_{M}\delta ^{\tilde{\beta}}_{N}\delta(z)
\nonumber
\\
&+&\sqrt{-\tilde{g_{5}}}V_{4}(y)\tilde{g}_{\tilde{\alpha}\tilde{\beta}}\delta ^{\tilde{\alpha}}_{M}\delta ^{\tilde{\beta}}_{N}\delta(z-\pi)
\end{eqnarray}
In the above expression $M$, $N$ are bulk indices,
$\alpha$, $\beta$ run over the usual four spacetime
coordinates given by $x^{\mu}$. The quantities $g$
and $\tilde{g}$ are the metric in $y=textrm{constant}$ and $z=\textrm{constant}$ hypersurfaces, respectively.
The line element derived from the above Einstein's equation turns out to be \cite{Sengupta}
\begin{equation}\label{bm4}
ds^{2}=\frac{\cosh^{2}(kz)}{\cosh^{2}(k\pi)}\left[exp(-2c\vert y \vert)\eta _{\mu \nu}dx^{\mu}dx^{\nu}+R_{y}^{2}dy^{2}\right]+r_{z}^{2}dz^{2}.
\end{equation}
In the above line element we have the following
identification for the constants  $k$ and $c$ given by
\begin{eqnarray}\label{bm5}
\left.\begin{array}{c}

c\equiv \frac{R_{y}k}{r_{z}\cosh(k\pi)}\\

k\equiv r_{z}\sqrt{\frac{-\Lambda _{6}}{10M^{4}}}.

\end{array} \right.
\end{eqnarray}
The boundary terms lead to the brane tensions and using
the Einstein's equation across the two boundaries
at $y=0$, $y=\pi$, respectively, thus we readily obtain
\begin{equation}\label{bm6}
V_{1}(z)=-V_{2}(z)=8M^{2}\sqrt{\frac{-\Lambda _{6}}{10}}\textrm{sech}(kz).
\end{equation}
Thus the two 4-branes situated at $y=0$ and $y=\pi$
would have a $z$-dependent brane tension. The fact
that the two tensions are equal and opposite is
reminiscent of the original RS-form. Similarly we get the brane tensions for other two 4-branes as
\begin{equation}\label{bm7}
V_{3}(y)=0;  V_{4}(y)=-\frac{8M^{4}k}{r_{z}}\tanh(k\pi).
\end{equation}
Here $V_{3,4}$ were introduced to account orbifolding
along the $z$-direction and with $g_{zz}$ being a constant,
the resulting hypersurface should have only a constant
energy density. The fact that $g_{yy}$ is dependent on
the coordinate $z$ makes the two hypersurfaces for
y orbifolding to have a z-dependent energy density.

The 3-brane located at $(y=0,z=\pi)$ suffers no warping
and can be identified with the Planck brane. The other
three can be valid choices for Standard Model (visible) brane.
However, if we assume that there is no brane having lower energy
scale than ours, we are forced to identify the SM brane to be
located at $(y=\pi ,z=0)$. The suppression factor on the TeV brane can be given by
\begin{equation}\label{branem01}
f=\frac{\exp(-c\pi)}{\cosh (k\pi)}.
\end{equation}
The desired suppression of $10^{-16}$ on the TeV brane can be
obtained by choosing different combinations of the parameters $c$
and $k$. However we also have an extra relation as presented in
Eq. (\ref{bm5}), which shows that in order to avoid large
hierarchy between the two moduli $R_{y}$ and $r_{z}$, either of
the two parameters $c$ and $k$ must be large and the other should
be small. For example, we can easily assume $c\sim 11.4$ and
$k\sim 0.1$. However a small hierarchy also exists in the original
RS model, where there exists an one order of magnitude hierarchy
between $r$ and $k$, satisfying Planck-to-TeV scale warping by
$kr\sim 11.5$. A natural question that arises with this discussion is
whether stabilization of these moduli to the desired values is
possible. For the five-dimensional RS model this has been shown in
\cite{Wise1} by introducing a bulk stabilizing scalar field and
tuning the VEV of the scalar field at the boundaries. The modulus
in the theory is stabilized near Planck length without any fine-tuning.

In this case as well we can adopt a similar procedure by
introducing a bulk stabilizing scalar field. Again choosing
appropriate VEV at the boundary, we can stabilize $R_{y}$ and
$r_{z}$ to desired values \cite{Bala05}. In our six-dimensional
braneworld scenario with $y$ and $z$ dependence, the action for
the scalar field can be expressed such that
\begin{eqnarray}
S&=&\int d^{4}xdydz\sqrt{-g_{6}}\left(\frac{1}{2}\partial _{M}\phi \partial ^{M}\phi -V(\phi) \right)
\nonumber
\\
&+&\sum _{i,j=1}^{2}\int d^{4}x \sqrt{-g_{ij}}\lambda _{ij}(\phi)\left(\phi ^{2}-v_{ij}^{2} \right)^{2}
\delta (y-y_{i})\delta (z-z_{j}),
\end{eqnarray}
where the coupling parameters, $\lambda _{ij}$ tend to infinity
as the scalar field approach to the following values,
$\phi(0,0)=v_{0}$, $\phi(0,\pi)=v_{1}$, $\phi(\pi,0)=v_{3}$ and
$\phi(\pi,\pi)=v_{4}$. We take $V(\phi) = m^2\phi^2$.
Now following Ref. \cite{Wise1,srikanth} we can obtain the equation of
motion in the separable form as
\begin{eqnarray}
\psi ''(y)-4c\psi '(y)=p\psi (y)
\nonumber
\\
\frac{b^{2}R_{y}^{2}}{r_{z}^{2}}\left[\ddot{\chi}(z)+\frac{5\dot{b}}{b}\dot{\chi}(z) \right]=
\left(R_{y}^{2}b^{2}m^{2}-p\right)\chi (z),
\end{eqnarray}
where we have made the following decomposition, $\phi (y,z)=\psi
(y)\chi (z)$ and $p$ is the separability constant. Also in the
above expression prime denote differentiation with respect to $y$
and dot denotes differentiation with respect to z. Finally, the
above equations with appropriate boundary condition
\cite{srikanth} can be solved, which when substituted into the
action leads to an effective potential for the moduli as
\begin{eqnarray}
V_{eff}&=&\pi v_{2}^{2}\Big[\frac{(1-2v+v^{2})}{2k\nu \pi}+\frac{k}{12\nu}
\Big((1+2\alpha -8\nu +2\nu ^{2})
\nonumber
\\
&+&v(22+2\alpha -8\nu +2\nu ^{2})+(1+2\alpha -8\nu +2\nu ^{2})\Big) \Big],
\end{eqnarray}
where we have used the following shorthand notations,
$v=v_{1}/v_{2}$, $\alpha =-10m^{2}M^{4}/\Lambda _{6}$ and $\nu
=\sqrt{4+\frac{p}{c^{2}}}$. Then solving the equations, $\partial
_{\nu}V_{eff}=0$ and $\partial _{k}V_{eff}=0$ and then through the
second derivatives with respect to $\nu$ and $k$ we can arrive at
the minimum values of $c$ and $k$. As an illustrative example we
can start with $v=0.43$ and $p\sim 1$, leading to $c\sim 11.24$
and $k\sim 0.422$. Note that these values can resolve the gauge
hierarchy problem. Thus along this line any higher-dimensional
braneworld models can have their moduli stabilized. From now on
we shall assume that such a stabilization has been performed and
all the moduli hence forth will have those stabilized values. In
this analysis, following the stabilizing bulk scalar model, we
have assumed that the backreaction of the bulk stabilizing field
is negligibly small. Moreover from the action of the bulk
stabilizing scalar it may be observed that at the boundaries the
stabilizing scalar tends to their VEVs $v_{ij}$ when the coupling
$\lambda_{ij}(\phi)$ tends to infinity. This is exactly similar to
the five-dimensional counter part of the Goldberger-Wise calculation
of modulus stabilization. As a result at the boundaries, the
stabilizing scalar is frozen to different values $v_{ij}$ and
hence does not contribute to the dynamics of the model.

\section{Bulk Field in Six-Dimensional Doubly Warped Spacetime}\label{bmbulk}

In this section we carry out the Kaluza-Klein decomposition
of a nongravitational bulk scalar field propagating in the
spacetime described by Eq. ($\ref{bm4}$) in the spirit
of the work \cite{Wise2} with bulk scalar field. We find that in these
multiply warped spacetime the SM brane contains larger number
of TeV scale scalar KK modes than the five-dimensional RS model. This has
significant phenomenological consequences \cite{Sen}.
We consider a free scalar field in the bulk for which the action is given by
\begin{equation}\label{bm8}
S=\frac{1}{2}\int d^{4}x\int dy \int dz\sqrt{-G}
\left(G^{AB}\partial _{A}\Phi \partial _{B}\Phi +m^{2}\Phi ^{2}\right),
\end{equation}
where $G_{AB}$ with $A,B=\mu ,y,z$ is given by Eq. ($\ref{bm4}$),
and m is of order of $M_{pl}$. After an integration by parts, this can be written as
\begin{eqnarray}\label{bm9}
S&=&\frac{1}{2}\int d^{4}x\int dy \int dz \Big[R_{y}r_{z}e^{-2\sigma}\frac{\cosh^{3}(kz)}{\cosh^{3}(k\pi)} \eta _{\mu \nu}\partial _{\mu}
\Phi \partial _{\nu}\Phi
\nonumber
\\
&+&R_{y}r_{z}e^{-4\sigma}\frac{\cosh^{5}(kz)}{\cosh^{5}(k\pi)}m^{2}\Phi _{2}
\nonumber
\\
&-&\frac{r_{z}}{R_{y}}\Phi \partial _{y} \left(e^{-4\sigma}\frac{\cosh^{3}(kz)}{\cosh^{3}(k\pi)}\partial _{y}\Phi \right)
\nonumber
\\
&-&\frac{R_{y}}{r_{z}}\Phi \partial _{z} \left(e^{-4\sigma}\frac{\cosh^{5}(kz)}{\cosh^{5}(k\pi)}\partial _{z}\Phi \right)\Big],
\end{eqnarray}
where $\sigma = c\vert y\vert$. Now we make the following substitution for KK decomposition,
\begin{equation}\label{bm10}
\Phi(x,y,z)=\sum _{n,m}\phi _{nm}(x)\frac{\alpha _{n}(y)}{\sqrt{R_{y}}}\frac{\beta _{m}(z)}{\sqrt{r_{z}}}.
\end{equation}
The following normalization conditions are imposed on the fields
$\alpha$ and $\beta$,
\begin{equation}\label{bm11}
\int dy e^{-2\sigma}\alpha _{n}\alpha _{m} = \delta _{nm}
\end{equation}
\begin{equation}\label{bm12}
\int dz \frac{\cosh^{3}(kz)}{\cosh^{3}(k\pi)} \beta _{p}\beta _{q} = \delta _{pq}.
\end{equation}
The differential equation satisfied by the function $\alpha _{n}(y)$ is
\begin{equation}\label{bm13}
-\frac{1}{R_{y}^{2}}\frac{d}{dy}\left(e^{-4\sigma}\frac{d\alpha _{m}}{dy}\right)=A_{m}^{2}e^{-2\sigma}\alpha _{m},
\end{equation}
where $A_{m}$ stands for KK mode mass eigenvalue.
The above differential equation can be further simplified and cast to the following form,
\begin{equation}\label{bm14}
\frac{d^{2}\alpha _{m}}{dy^{2}}-4c\frac{d\alpha _{m}}{dy}+A_{m}^{2}R_{y}^{2}e^{2\sigma }\alpha _{m}=0.
\end{equation}
The above equation can be solved in terms of Bessel functions of first and second order as
\begin{equation}\label{bm14a}
\alpha _{m}=\frac{e^{2\sigma}}{N_{m}}\left[J_{2}\left(\frac{A_{m}e^{\sigma}R_{y}}{c}\right)+
b_{m}Y_{2}\left(\frac{A_{m}e^{\sigma}R_{y}}{c} \right) \right],
\end{equation}
with $N_{m}$ representing an overall normalization.
\begin{figure}

\includegraphics[height=3in,width=4in]{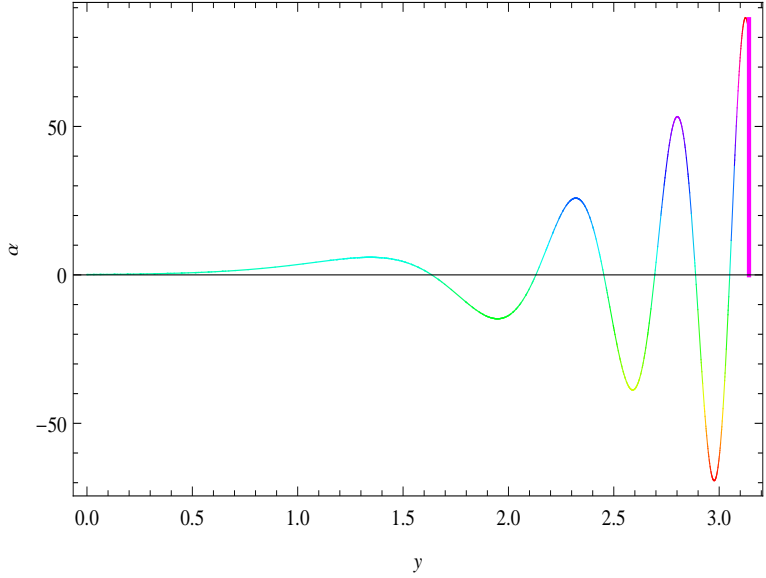}

\vskip 3mm
\caption {The figure shows variation of the quantity
$\alpha _{m}$ with extra-dimension parameter $y$. The vertical
line represents the $y=\pi$ line showing the fact that
the quantity $\alpha _{m}$ is maximum at $y=\pi$, the position
of TeV brane.} \label{fig1}

\end{figure}
Now we can proceed much further. The mass modes determined by $A_{m}$ must be real.
This reality condition imposed on the mass modes requires the differential operator on the left
hand side of Eq. (\ref{bm14a}) to be self-adjoint. This self-adjointness imply that derivatives of
$\alpha _{m}(y)$ should be continuous at the orbifold fixed points.
These gives two conditions on the parameters $A_{m}$
and $b_{m}$, expressed as,
\begin{eqnarray}\label{bmnew1}
b_{m}&=&-\frac{2J_{2}\left(\frac{A_{m}R_{y}}{c}\right)+\frac{A_{m}R_{y}}{c}J'_{2}\left(\frac{A_{m}R_{y}}{c}\right)}
{2Y_{2}\left(\frac{A_{m}R_{y}}{c}\right)+\frac{A_{m}R_{y}}{c}Y'_{2}\left(\frac{A_{m}R_{y}}{c}\right)}\\
0&=&\left[2J_{2}\left(x_{m}\right)+x_{m}J'_{2}\left(x_{m}\right)\right]
\left[2Y_{2}(x_{m}e^{-c\pi})+x_{m}e^{-c\pi}Y'_{2}\left(x_{m}e^{-c\pi}\right)\right]
\nonumber
\\
&-&\left[2Y_{2}\left(x_{m}\right)+x_{m}Y'_{2}\left(x_{m}\right)\right]
\left[2J_{2}(x_{m}e^{-c\pi})+x_{m}e^{-c\pi}J'_{2}\left(x_{m}e^{-c\pi}\right)\right]
\end{eqnarray}
where $x_{m}=A_{m}e^{c\pi}R_{y}/c$. Since to make Planck scale
down to TeV scale we should have $e^{c\pi}\gg 1$. Then the above equation
reduces to the following form,
$2J_{2}\left(x_{m}\right)+x_{m}J'_{2}\left(x_{m}\right)=0$.
Then for light mode masses we have $x_{1}$ to be order
of unity \cite{Wise2}. This keeps $A_{m}R_{y}/c$ also order
of unity. Then from Eq. (\ref{bm14a}) as well as from Fig.\ref{fig1}
we observe that the
modes $\alpha _{m}(y)$ are larger near the 3-brane at $y=\pi$,
which makes these low-mass Kaluza-Klein modes to be found preferentially
near the $y=\pi$ region (see Fig.\ref{fig1}). Thus, with the TeV brane being
situated at $y=\pi$, we observe that the low-mass KK modes
are exponentially suppressed and
hence confined to the TeV brane.
Also for these low-lying KK mass modes the coefficient
$b_{m}$ is of the order of $e^{-4c\pi}$, which shows
that we can ignore the $Y_{2}(y_{m})$
part compared to $J_{2}(y_{m})$, while performing
integrals involving $\alpha _{m}$.

\begin{figure}

\includegraphics[height=3in,width=4in]{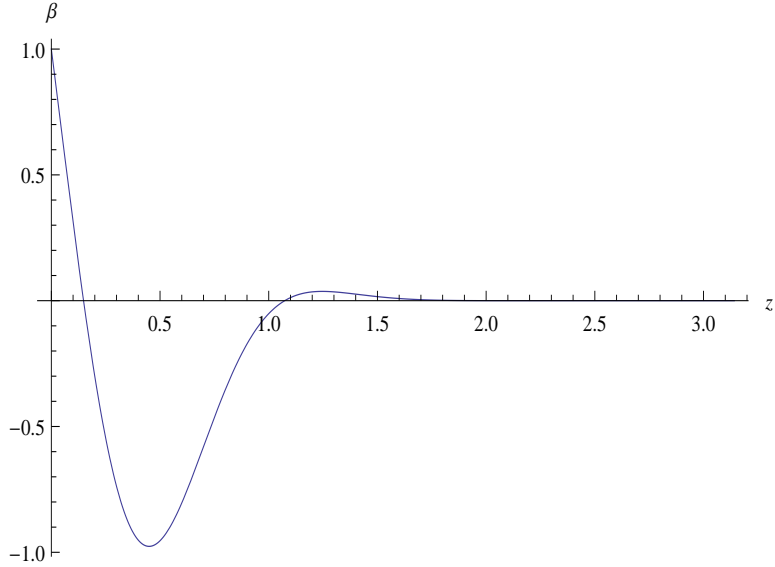}

\vskip 3mm
\caption{The figure shows variation of the quantity $\beta _{n}$ with extra-dimension parameter $z$. The
graph clearly shows the fact that the quantity $\beta _{n}$ is maximum at $z=0$, the position
of TeV brane.} \label{fig2}

\end{figure}

Similar analysis for $\beta _{m}$ yields
\begin{equation}\label{bm15}
\frac{d^{2}\beta _{m}}{dz^{2}}+5k\tanh(kz)\frac{d\beta _{m}}{dz}+r_{z}^{2}B_{m}^{2}\frac{\cosh^{2}(k\pi)}{\cosh^{2}(kz)}\beta _{m}=0.
\end{equation}
The solution, apart from an overall normalization, can be expressed as
\begin{eqnarray}\label{bm15a}
\beta _{m}(z)&=&exp\left[-\frac{5}{2}k^{2}z^{2}\right] H_{\sqrt{5/2}kz}\left(\frac{-10k^{2}+B_{m}^{2}r_{z}^{2}(1+\cosh(2k\pi))}{10k^{2}}\right)
\nonumber
\\
&+&E_{m}exp\left[-\frac{5}{2}k^{2}z^{2}\right]~_{1}F_{1}
\left(-\frac{-10k^{2}+B_{m}^{2}r_{z}^{2}(1+\cosh(2k\pi))}{10k^{2}},\frac{1}{2},\frac{5k^{2}z^{2}}{2}\right),
\end{eqnarray}
where $_{1}F_{1}$ is the Kummer confluent hypergeometric function
and $H_{n}$ is the Hermite polynomial of degree n. Then from
Fig.\ref{fig2} we observe that this function is also maximum at
$z=0$. Hence the low-lying KK mass modes are confined to the TeV
brane located at $z=0$. From the behavior of both $\alpha _{m}$
and $\beta _{n}$ we find that all the low-lying KK mass modes are
confined to $y=\pi ,z=0$ brane, i.e., the TeV brane. A possible
experimental signature of the bulk scalar KK modes can originate
via coupling of the bulk scalar to diHiggs in the form $\Phi
(x)h^{2}(x)$. For $m_{\Phi}\sim m_{h}$ the dominant decay channels
are $gg$ and $b\bar{b}$ which leads to multijets as final product
which though may be difficult to differentiate from the QCD back
ground \cite{Cox2014,Djouadi2008,Csaki2007}. Also when the mass of
bulk scalar is in the range of $250$ to $350$ GeV then enhanced
production of $\Phi \rightarrow hh$ occurs. Moreover for bulk
scalar mass in the range $160$ to $250$ GeV we have a relatively
larger cross sections for the diphoton channel. In this region due
to small mixing the branching ratio is dominated by $gg$ and
$b\bar{b}$. The diphoton channel is  a very  promising search
channel as branching ratio remains more or less at constant level
even up to $t\bar{t}$ threshold \cite{CMS2013a,CMS2013b}. This
might become possible if the LHC runs extends the diphoton
searches
for invariant masses above existing $m_{\gamma \gamma} =150$GeV.\\

To determine the parameters of the solution we proceed as follows:
we want $B_{m}$ to be real, as it
appears in the mass modes. Thus self-adjointness also
applies in this case. This implies that derivatives of
$\beta _{m}$ to be continuous around the orbifold fixed
points. At $z=0$, this is trivially satisfied irrespective
of the quantity $E_{m}$. However at $z=\pi$ all the terms
are suppressed by $\exp(-5k^{2}z^{2}/2)$, thus the self-adjointness there leads to
\begin{eqnarray}
E_{m}&=&-\frac{H_{\sqrt{5/2}k\pi}\left(a\right)}{{1}F_{1}
\left(-a,\frac{1}{2},\frac{5k^{2}z^{2}}{2}\right)+2a~_{1}F_{1}\left(a+1,\frac{3}{2},-\frac{5}{2}k^{2}z^{2} \right)}\\
a&=&\frac{-10k^{2}+B_{m}^{2}r_{z}^{2}(1+\cosh(2k\pi))}{10k^{2}}.
\end{eqnarray}
At large values of $z$, confluent hypergeometric
series have a large value. Being in the denominator the term can be neglected for
practical purposes.
Using the above equations we readily obtain the following action
for the field $\phi (x)$ as,
\begin{eqnarray}\label{bm16}
S&=&\frac{1}{2}\int d^{4}x \Big(\sum _{n,m}\eta ^{\mu \nu}\partial _{\mu}\phi _{nm}\partial _{\nu}\phi _{nm}
+\sum _{n,m,p,q} M_{nmpq}\phi _{nm}\phi _{pq}\Big)
\\
M_{nmpq}&=&\left\lbrace A_{n}^{2}\delta _{np}\delta _{mq}+B_{n}^{2}\delta _{np}P_{mq}+m^{2}P_{np}Q_{mq}\right\rbrace
\end{eqnarray}
where we have the following expression for the element $Q_{nm}$,
\begin{equation}\label{bm17}
Q_{nm}=\int dz \frac{\cosh^{5}(kz)}{\cosh^{5}(k\pi)} \beta _{n}\beta _{m}
\end{equation}
and $P_{mn}$ as,
\begin{equation}\label{bm17a}
P_{nm}=\int dy e^{-4\sigma}\alpha _{n}\alpha _{m}
\end{equation}
Now from the previous discussion we have the solution for these two sets
of functions $\alpha _{n}(y)$ and $\beta _{n}(z)$, which can be used
to determine $Q_{nm}$ and $P_{nm}$ in order to obtain the masses of
the KK modes by evaluating the quantity $M_{nmpq}$.
In contrast to the five-dimensional situation (see \cite{Wise2})
where the masses of the bulk fields appear as a diagonalized mass matrix,
in this case the bulk field $\Phi (x,y,z)$ manifests itself to some
four-dimensional observer as an infinite KK tower with mass being
determined by the the quantity $M_{nmpq}$ such that a scalar
$\phi _{nm}$ has a mass $M_{nmnm}$ after an appropriate diagonalization procedure.

The solution for $\alpha _{n}(y)$ is presented in equation
(\ref{bm14a}). A similar solution was obtained by Wise et. al (see \cite{Wise2})
except for the fact that we have Bessel functions of second order.
Following their discussion we can argue in a similar manner that the lightest
KK modes have mass parameter $A_{m}$ suppressed exponentially with
respecanct to the the scale $m$ appearing in Eq. (\ref{bm8}).
Since we have taken $m$ to be order of Planck scale and $c$ to be
around $12$, by stabilization these mass modes $A_{m}$ are in the TeV range.
Also from the solution we could observe that the modes $\alpha _{n}(y)$
are larger in the region $y=\pi$. This has been explained earlier through
graphical presentation of the function $\alpha _{m}$.

The solution for $\beta _{n}(z)$ has been presented in Eq.
(\ref{bm15a}). Though we have argued following the graphical presentation
of the function $\beta _{n}$, we now provide a theoretical motivation for our
above-mentioned results. The solution has an overall factor of
$exp\left[-\frac{5}{2}k^{2}z^{2}\right]$ and we see that the
solution has maximum value around $z=0$. Hence the bulk field
being a product of these two functions $\alpha _{m}(y)$ and $\beta
_{n}(z)$ as shown in Eq. (\ref{bm9}) has mass parameter in
the TeV range and has maximum value around $(y=\pi ,z=0)$. Now
from Sec. \ref{bmsix} this is precisely the SM brane. Hence, the
bulk field has a maximum in the SM brane; i.e., the KK modes are
most likely to be found in that region where $A_{m}$ and $B_{m}$
are in the TeV range. This sets the stage for KK excitations to
have TeV scale mass splitting on the SM brane.

Now we would like to compute some low-lying KK mode
masses numerically. For that we need to fix some parameters,
$k$, $c$, $r_{z}$ and the bulk mass of the scalar field $m$.
We shall take the bulk mass to be in Planck scale. Then we
can determine the remaining parameters, by making the following
demands: (a) if we have a gauge boson field in this multiply
warped scenario, its lowest massive KK modes should lead to
$W$ and $Z$ boson masses, (b) the suppression $f$ as presented
in Eq. (\ref{branem01}) should be $\sim 10^{-16}$, and finally
(c) the hierarchy between $R_{y}$ and $r_{z}$ should be small.
The KK mode of the gauge boson in this multiply warped spacetime
can be obtained from Ref. \cite{Ashmita11}.

This desired mass
for $W$ and $Z$ boson $\sim 100$GeV can be obtained with
$f\sim 10^{-16}$ and $\frac{1}{r_{z}}=7\times 10^{17}$GeV,
about 14 times smaller compared to Planck scale. The other
parameters $k$ and $c$ can be determined using small hierarchy
between $R_{y}$ and $r_{z}$ along with desired warping of
$f\sim 10^{-16}$. This finally leads to, the following estimation:
$k=0.25$, $c=11.52$ and the ratio between moduli being
$\frac{R_{y}}{r_{z}}=61$. The suppression factor turns out
to be $f=1.45 \times 10^{-16}$. Thus we will calculate the
low-lying KK masses for our bulk scalar field with these sets of parameters (see Table.\ref{Tab1}).

We now present the self-interactions of the bulk scalar field.
From the four-dimensional point of view these self-interactions
can induce couplings between the KK modes. In this case
self-couplings of the light modes are suppressed by the warp factor
and hence if the Planck scale set the six-dimensional couplings,
the low-lying KK modes have TeV range self-interactions. We present
the interaction term in the action with coupling parameter $\lambda$ such that
\begin{equation}\label{bmm1}
S_{int}=\int d^{4}x \int _{-\pi}^{\pi}dy \int _{-\pi}^{\pi}dz \sqrt{G} \frac{\lambda}{M^{4m-6}}\Phi ^{2m},
\end{equation}
where the coupling $\lambda$ is of the order of unity. Then we can expand in modes and the self-interactions of light KK states become
\begin{equation}\label{bmm2}
S_{int}=\int d^{4}x \int _{-\pi}^{\pi}dy \int _{-\pi}^{\pi}dz R_{y}r_{z}e^{-4\sigma}
 \frac{\cosh^{5}(kz)}{\cosh^{5}(k\pi)}
\frac{\lambda}{M^{4m-6}}\phi _{pq} ^{2m}\left(\frac{\alpha _{p}}{\sqrt{R_{y}}}\frac{\beta _{q}}{\sqrt{r_{z}}}\right)^{2m}.
\end{equation}
Thus the effective four dimensional coupling constants are
\begin{equation}\label{bmm3}
\lambda _{eff}=\frac{4\lambda}{(MR_{y})^{m-1}(Mr_{z})^{m-1}M^{2m-4}}\int _{0}^{\pi}dy e^{-4\sigma}\alpha _{p}^{2m}
\int _{0}^{\pi}dz \frac{\cosh ^{5}(kz)}{\cosh ^{5}(k\pi)}\beta _{q}^{2m},
\end{equation}
which reduces to,
\begin{equation}\label{bmm4}
\lambda _{eff}\simeq 4\lambda \left(\frac{c}{MR_{y}}\right)^{m-1}\left(\frac{1}{Mr_{z}}\right)^{m-1}
\left(Me^{-c\pi}\frac{1}{\cosh(k\pi)}\right)^{4-2m}\int _{0}^{1}r^{4m-5}dr
\left[\frac{J_{2}\left(\frac{A_{p}e^{\sigma}}{k}r\right)}{A_{p}}\right]^{2m}\int _{0}^{\pi}(\beta _{q})^{2m}dz
\end{equation}
in the large $kR_{y}$ and $kr_{z}$ limit. Hence we observe that the
relevant scale for four-dimensional physics is not the scale set
by Planck scale, i.e., $M$, but this is
$Me^{-c\pi}\frac{1}{cosh~k\pi}$. Hence the KK reduction has lead
the couplings from Planck scale to the TeV scale by the warp
factor on the SM brane located at $(y=\pi ,z=0)$.
\begin{table}
\begin{center}

\caption{The masses of the KK modes of the scalar field are given
in GeV units. We have chosen the following values,
$\frac{1}{r_{z}}=7\times 10^{17}$ GeV, $k=0.25$, $c=11.52$. Some
representative masses of low-lying KK modes are given.}
\label{Tab1} \centering

\begin{tabular}{|c|c|c|c|}

\hline
\hline

$m_{1111}=99.513$
&
$m_{1212}=99.651$
&
$m_{1313}=99.709$
&
$m_{1414}=99.743$\\

$m_{2121}=178.614$
&
$m_{2222}=178.866$
&
$m_{2323}=178.965$
&
$m_{2424}=179.026$\\

$m_{3131}=257.714$
&
$m_{3232}=258.069$
&
$m_{3333}=258.228$
&
$m_{3434}=258.309$\\

$m_{4444}=337.592$
&
$m_{5555}=416.957$
&
$m_{6666}=501.445$
&
$m_{7777}=583.371$\\

\hline
\hline

\end{tabular}
\end{center}
\end{table}

From the above discussion we now try to obtain some bounds on the
parameters in our model, e.g., $R_{y}$, $r_{z}$ from the requirement
of precision electroweak test. For that purpose we can use the
same setup and put a bulk gauge boson whose KK modes can be
detected in precision electroweak tests. We define a quantity
denoted by
\begin{equation}
V=\sum _{n=1}^{\infty}\left( \frac{g_{n}^{2}}{g_{0}^{2}}\frac{M_{W}^{2}}{M_{n}^{2}}\right),
\end{equation}
where $M_{W}$ is the mass of $W$ gauge boson and $M_{n}$ is the
mass of higher KK modes of the bulk gauge boson and  $g_{0}$ is
the effective four-dimensional gauge coupling along with $g_{n}$
to be the gauge couplings for higher KK modes. Then from Ref.
\cite{Rizzo00} we could argue that for precision electroweak test
we should have $V<0.0013$ with $95$ percent confidence level. From
this result we can get the following bounds on the parameters of
this model, $1/R_{y}<5.95 \times 10^{17}$GeV. This leads to a
bound on $1/r_{z}$ as well by assuming a small hierarchy between
the two moduli as, $1/r_{z}<3.63\times 10^{19}$GeV. From Ref.
\cite{Ashmita11} it can be easily verified that this bound is
respected by gauge couplings and KK mode masses. Thus these
multiply warped models indeed satisfy precision electroweak tests.

\section{Seven-and-Higher-Dimensional Spacetime with Multiple Warping}\label{bmseven}

With an aim to arrive at a generic result we shall now try to
extend our analysis with one more extra dimension. For that
purpose we start with a seven-dimensional spacetime where the
spacelike dimensions are successively warped. In other words the
manifold of interest could be given by $\left[\left\lbrace
M^{(1,3)}\times \left[ S^{1}/Z_{2}\right]\right\rbrace \times
S^{1}/Z_{2}\right]\times S^{1}/Z_{2}$. Then the total brane-bulk
action can be given by
\begin{eqnarray}\label{bm19}
S&=&S_{7}+S_{6}+S_{5}+S_{4}
\\
S_{7}&=&\int d^{4}xdydzdw \sqrt{-g_{7}}\left(R_{7}-\Lambda _{7}\right)
\\
S_{6}&=&\int d^{4}xdydzdw \left[V_{1}\delta(w)+V_{2}\delta(w-\pi)\right]
\nonumber
\\
&+&\int d^{4}xdydzdw \left[V_{3}\delta(z)+V_{4}\delta(z-\pi)\right]
\nonumber
\\
&+&\int d^{4}xdydzdw \left[V_{5}\delta(y)+V_{6}\delta(y-\pi)\right],
\end{eqnarray}
with appropriate actions ($S_{5}$) for 12 possible 4-branes at the edges
$(z,w)=(0,0),(0,\pi),(\pi ,0),(\pi ,\pi)$, $(z,y)=(0,0),(0,\pi),(\pi ,0),(\pi ,\pi)$ and $(y,w)=(0,0),(0,\pi),(\pi ,0),(\pi ,\pi)$.
We also have eight possible 3-branes at the corners
$(y,z,w)=(0,0,0),(0,0,\pi),(0,\pi ,0),(\pi ,0,0),(\pi ,\pi ,0),(0,\pi ,\pi),(\pi ,0,\pi),(\pi ,\pi ,\pi)$. By natural extension of the method as
illustrated in the previous section we get the line element and other parameters such that \cite{Sengupta},
\begin{eqnarray}\label{bm20}
ds^{2}&=&\frac{\cosh^{2}(\ell w)}{\cosh^{2}(\ell \pi)}\left\lbrace \frac{\cosh^{2}(kz)}{\cosh^{2}k\pi}
\left[exp(-2c|y|)\eta _{\mu \nu}dx^{\mu}dx^{\nu}+
R_{y}^{2}dy^{2}\right]+r_{z}^{2}dz^{2}\right\rbrace +\Re _{w}^{2}dw^{2}
\nonumber
\\
\ell ^{2}&=&\frac{-\Lambda _{7}\Re _{w}^{2}}{15}
\nonumber
\\
k&=& \frac{\ell r_{z}}{\Re _{w}\cosh(\ell \pi)}
\nonumber
\\
c&=&\frac{\ell R_{y}}{\Re _{w}\cosh(k\pi)cosh(\ell \pi)}=\frac{kR_{y}}{r_{z}\cosh(k\pi)}
\end{eqnarray}
It may be of interest that the 5-brane at $w=\pi$ does not
represent a flat metric (y and z dependencies). In order to obtain
substantial warping along the $w$ direction (from $w=\pi$ to
$w=0$), one need to make $\ell \pi$ substantial (same order of
magnitude as RS scenario). The seven-dimensional or triply warped
model has a structure analogous to that of six-dimensional one,
not only in the, form of functional dependence but also on the
nature of warping. This method can easily be extended to even
higher dimensions. Also note that the orbifolding requires branes
situated at edges of n-dimensional hypercube with 3-branes at the
corners. If one of the direction suffers a large warping any other
direction should have small warping so that there is no large
hierarchy coming from the moduli. In this case also we have
several candidates for our SM brane. However applying the fact
that no brane should have less energy than ours, leads to $(y=\pi
,z=0,w=0)$ to be SM brane.

\section{Bulk Fields in Seven-and-Higher-Dimensional Spacetime}\label{bmbulk7}

Following the methods of previous sections, we shall carry out the
Kaluza-Klein decomposition of a bulk scalar field propagating
in the spacetime given by Eq. ($\ref{bm20}$). As in the
previous section in this case as well we can write the bulk scalar
field in terms of product of four functions. By making KK
decomposition we again end up with KK mass modes having TeV scale masses and
splittings. The action for the bulk scalar field in this seven-dimensional spacetime can be given as
\begin{equation}\label{bm21}
S=\frac{1}{2}\int d^{4}x\int dy\int dz\int dw \sqrt{-G}\left[G_{AB}\partial ^{A}\Phi \partial ^{B}\Phi +m^{2}\Phi ^{2}\right].
\end{equation}
From the line element as given by Eq. ($\ref{bm19}$), we readily obtain the following form for the action
\begin{eqnarray}\label{bm22}
S&=&\frac{1}{2}\int d^{4}x\int dy\int dz\int dw
\Big[R_{y}r_{z}\Re _{w}e^{-2\sigma}
\frac{\cosh^{4}(\ell w)}{\cosh^{4}(\ell \pi)}\frac{\cosh^{3}(kz)}{\cosh^{3}(k\pi)}\eta _{\mu \nu}\partial _{\mu}\Phi \partial _{\nu}\Phi
\nonumber
\\
&+&\frac{1}{2}\frac{\Re _{w}r_{z}}{R_{y}}\frac{\cosh^{4}(\ell w)}{\cosh^{4}(\ell \pi)}
\frac{\cosh^{3}(kz)}{\cosh^{3}(k\pi)}e^{-4\sigma}(\partial _{y}\Phi)^{2}
\nonumber
\\
&+&\frac{1}{2}\frac{\Re _{w}R_{y}}{r_{z}}\frac{\cosh^{4}(\ell w)}{\cosh^{4}(\ell \pi)}
\frac{\cosh^{5}(kz)}{\cosh^{5}(k\pi)}e^{-4\sigma}(\partial _{z}\Phi)^{2}
\nonumber
\\
&+&\frac{1}{2}\frac{r_{z}R_{y}}{\Re _{w}}\frac{\cosh^{6}(\ell w)}{\cosh^{6}(\ell \pi)}
\frac{\cosh^{5}(kz)}{\cosh^{5}(k\pi)}e^{-4\sigma}(\partial _{w}\Phi)^{2}
\nonumber
\\
&+&\frac{1}{2}m^{2}\Re _{w}r_{z}R_{y}\frac{\cosh^{6}(\ell w)}{\cosh^{6}(\ell \pi)}
\frac{\cosh^{5}(kz)}{\cosh^{5}(k\pi)}e^{-4\sigma}\Phi ^{2}\Big],
\end{eqnarray}
where $\sigma =c\vert y\vert$. We make the following substitution for the bulk field:
\begin{equation}\label{bm23}
\Phi =\sum _{pqr}\phi _{pqr}(x)\frac{\alpha _{p}}{\sqrt{R_{y}}}\frac{\beta _{q}}{\sqrt{r_{z}}}\frac{\gamma _{r}}{\sqrt{\Re _{w}}}
\end{equation}
We also impose the following normalization for the functions $\alpha _{p}$, $\beta _{q}$ and $\gamma _{r}$,
\begin{eqnarray}\label{bm24}
\int e^{-2}\alpha _{m}\alpha _{n}dy&=&\delta _{mn}
\\
\int \frac{\cosh^{3}(kz)}{\cosh^{3}(k\pi)}\beta _{m}\beta _{n}dz&=&\delta _{mn}
\\
\int \frac{\cosh^{4}(\ell w)}{\cosh^{4}(\ell \pi)}\gamma _{m}\gamma _{n}dw&=&\delta _{mn}.
\end{eqnarray}
Now applying integration by parts to the integral as presented in Eq. ($\ref{bm22}$) we readily obtain
\begin{eqnarray}\label{bm25}
S&=&\frac{1}{2}\int d^{4}x\int dy\int dz\int dw \Big \lbrace
\left[\sum_{pqrabc} e^{-2\sigma}\frac{\cosh^{4}(\ell w)}{\cosh^{4}(\ell \pi)}\frac{\cosh^{3}(kz)}{\cosh^{3}(k\pi)}
\left(\eta ^{\mu \nu}\partial _{\mu}\phi _{pqr} \partial _{\nu}\phi _{abc}\right)
\alpha _{p}\alpha _{a}\beta _{q}\beta _{b}\gamma _{r}\gamma _{c}\right]
\nonumber
\\
&-&\frac{1}{2}\frac{1}{R_{y}^{2}}\left[\sum_{pqrabc} \frac{\cosh^{4}(\ell w)}{\cosh^{4}(\ell \pi)}\frac{\cosh^{3}(kz)}{\cosh^{3}(k\pi)}
\phi _{pqr}\phi _{abc}\beta _{q}\beta _{b}\gamma _{r}\gamma _{c}\alpha _{p}
\partial _{y}\left(e^{-4\sigma}\partial _{y}\alpha _{a}\right)\right]
\nonumber
\\
&-&\frac{1}{2}\frac{1}{r_{z}^{2}}\left[\sum_{pqrabc} \frac{\cosh^{4}(\ell w)}{\cosh^{4}(\ell \pi)}e^{-4\sigma}
\phi _{pqr}\phi _{abc}\alpha _{p}\alpha _{a}\gamma _{r}\gamma _{c}\beta _{q}
\partial _{z}\left(\frac{\cosh^{5}(kz)}{\cosh^{5}(k\pi)}\partial _{z}\beta _{b}\right)\right]
\nonumber
\\
&-&\frac{1}{2}\frac{1}{\Re _{w}^{2}}\left[\sum \frac{\cosh^{5}(kz)}{\cosh^{5}(k\pi)}e^{-4\sigma}
\phi _{pqr}\psi _{abc}\alpha _{p}\alpha _{a}\beta _{q}\beta _{b}\gamma _{r}
\partial _{w}\left(\frac{\cosh^{6}(\ell w)}{\cosh^{6}(\ell \pi)}   \partial _{w}\gamma _{c}\right)\right]
\nonumber
\\
&+&\frac{1}{2}m^{2}\left[\sum \frac{\cosh^{6}(\ell w)}{\cosh^{6}(\ell \pi)}\frac{\cosh^{5}(kz)}{\cosh^{5}(k\pi)}e^{-4\sigma}
\phi _{pqr}\phi _{abc}\alpha _{p}\alpha _{a}\beta _{q}\beta _{b}\gamma _{r}\gamma _{c}\right] \Big \rbrace.
\end{eqnarray}
Then we make the following choice for the differential equations satisfied by the functions $\alpha _{n}$, $\beta _{n}$ and $\gamma _{n}$,
\begin{eqnarray}
-\frac{1}{R_{y}^{2}}\partial _{y}\left(e^{-4\sigma}\partial _{y}\alpha _{n} \right)&=&A_{n}^{2}e^{-2\sigma}\alpha _{n}
\label{bmm26a}
\\
-\frac{1}{r_{z}^{2}}\partial _{z}\left(\frac{\cosh^{5}(kz)}{\cosh^{5}(k\pi)}\partial _{z}\beta _{n}\right)&=&
B_{n}^{2}\frac{\cosh^{3}(kz)}{\cosh^{3}(k\pi)}\beta _{n}
\label{bmm26b}
\\
-\frac{1}{\Re _{w}^{2}}\partial _{w}\left(\frac{\cosh^{6}(\ell w)}{\cosh^{6}(\ell \pi)}   \partial _{w}\gamma _{n}\right)&=&
C_{n}^{2}\frac{\cosh^{4}(\ell w)}{\cosh^{4}(\ell \pi)}\gamma _{n}.
\label{bmm26c}
\end{eqnarray}
The first equation as presented in (\ref{bmm26a}) can be
solved and has an identical solution as that obtained in the previous section.
However, for convenience we rewrite the solution,
\begin{equation}\label{bm26a}
\alpha _{p}=\frac{e^{2\sigma}}{N_{p}}\left[J_{2}\left(\frac{A_{p}e^{\sigma}R_{y}}{c}\right)+
b_{p}Y_{2}\left(\frac{A_{p}e^{\sigma}R_{y}}{c} \right) \right]
\end{equation}
The second equation as given by Eq. (\ref{bmm26b})
has the same solution as presented in Eq. (\ref{bm15a}) but we rewrite it here,
\begin{eqnarray}\label{bm26b}
\beta _{q}(z)&=&exp\left[-\frac{5}{2}k^{2}z^{2}\right] H_{\sqrt{5/2}kz}\left(\frac{-10k^{2}+B_{q}^{2}r_{z}^{2}(1+\cosh(2k\pi))}{10k^{2}}\right)
\nonumber
\\
&+&E_{q}exp\left[-\frac{5}{2}k^{2}z^{2}\right]~_{1}F_{1}
\left(-\frac{-10k^{2}+B_{q}^{2}r_{z}^{2}(1+\cosh(2k\pi))}{10k^{2}},\frac{1}{2},\frac{5k^{2}z^{2}}{2}\right)
\end{eqnarray}
The third Eq. (\ref{bmm26c}) has the following solution along with an overall normalization,
\begin{eqnarray}\label{bm26c}
\gamma _{r}(w)&=&exp\left[-3\ell ^{2}w^{2}\right] H_{\sqrt{3}\ell w}
\left(\frac{-12\ell ^{2}+C_{r}^{2}\Re_{w}^{2}(1+\cosh(2\ell \pi))}{12\ell ^{2}}\right)
\nonumber
\\
&+&F_{r}exp\left[-3\ell ^{2}w^{2}\right]~_{1}F_{1}
\left(-\frac{-12\ell ^{2}+C_{r}^{2}\Re_{w}^{2}(1+\cosh(2\ell \pi))}{24\ell ^{2}},\frac{1}{2},3\ell ^{2}w^{2}\right)
\end{eqnarray}
Here also we have $J_{2}$ and $Y_{2}$ to be Bessel functions of first and second order respectively. 
Along with these $H_{n}$ represents Hermite polynomials and $_{1}F_{1}$ is the Kummer 
confluent hypergeometric series. The arbitrary constants $b_{p}$, $E_{q}$ and $F_{r}$ can be 
determined by the self-adjoint criteria and have the following expressions
\begin{eqnarray}\label{branem001}
b_{m}&=&-\frac{2J_{2}\left(\frac{A_{p}R_{y}}{c}\right)+\frac{A_{p}R_{y}}{c}J'_{2}\left(\frac{A_{p}R_{y}}{c}\right)}
{2Y_{2}\left(\frac{A_{p}R_{y}}{c}\right)+\frac{A_{p}R_{y}}{c}Y'_{2}\left(\frac{A_{m}R_{y}}{c}\right)}\\
E_{q}&=&-\frac{H_{\sqrt{5/2}k\pi}\left(a\right)}{{1}F_{1}
\left(-a,\frac{1}{2},\frac{5k^{2}z^{2}}{2}\right)+2a~_{1}F_{1}\left(a+1,\frac{3}{2},-\frac{5}{2}k^{2}z^{2} \right)}\\
a&=&\frac{-10k^{2}+B_{q}^{2}r_{z}^{2}(1+\cosh(2k\pi))}{10k^{2}}
\nonumber
\\
F_{r}&=&-\frac{H_{\sqrt{5/2}k\pi}\left(b\right)}{{1}F_{1}
\left(-b,\frac{1}{2},\frac{5k^{2}z^{2}}{2}\right)+2b~_{1}F_{1}\left(b+1,\frac{3}{2},-\frac{5}{2}k^{2}z^{2} \right)}\\
b&=&\frac{-10k^{2}+C_{r}^{2}\Re _{w}^{2}(1+\cosh(2k\pi))}{10k^{2}}.
\nonumber
\end{eqnarray}
Hence our final expression for the action is given by
\begin{eqnarray}
S&=&\frac{1}{2}\int d^{4}x \Big[\sum _{pqr} \eta ^{\mu \nu}\partial _{\mu}\phi _{pqr}\partial _{\nu}\phi _{pqr}
+\sum _{abcpqr} M_{pqrabc} \phi _{pqr}\phi _{abc}\Big]
\label{bm27a}
\\
M_{pqrabc}&=&\left\lbrace A_{p}^{2}\delta _{pa}\delta _{qb}\delta _{rc}+B_{p}^{2}P_{rc}\delta _{pa}\delta _{qb}
+C_{p}^{2}P_{qb}Q_{rc}\delta _{pa}+~m^{2}P_{pa}Q_{qb}R_{rc}\right\rbrace,
\label{bm27b}
\end{eqnarray}
where we have defined the following quantities,
\begin{eqnarray}
P_{mn}=\int dy e^{-4cy}\alpha _{n}\alpha _{m}
\label{bm28a}
\\
Q_{mn}=\int dz \frac{\cosh^{5}(kz)}{\cosh^{5}(k\pi)}\beta _{n}\beta _{m}
\label{bm28b}
\\
R_{mn}=\int dw \frac{\cosh^{6}(\ell w)}{\cosh^{6}(\ell \pi)} \gamma _{n}\gamma _{m}.
\label{bm28c}
\end{eqnarray}
Now from the previous discussion we can find the solution for
the three sets of functions $\alpha _{n}(y)$, $\beta _{n}(z)$ and $\gamma _{n}(w)$
which in turn determine $P_{mn}$, $Q_{mn}$ and $R_{mn}$.
Therefore from these three functions the explicit expression for
KK mass modes can be determined from the quantity $M_{pqrabc}$
as given by Eq. (\ref{bm27b}). In this case as well the
bulk field $\Phi (x,y,z)$ manifests itself to some four-dimensional 
observer as a scalar $\phi _{pqr}$ whose mass is determined
by Eqs. (\ref{bm27a}) and (\ref{bm27b}).

The solution for $\alpha _{n}(y)$ as presented in Eq.
(\ref{bm26a}) has the same nature as obtained by Wise \textit{et al.} (see \cite{Wise2}).
In this case as well lightest KK modes have mass modes
determined by $A_{m}$, suppressed exponentially with
respect to the the scaler mass $m$
which we have taken to be order of Planck scale. Thus
these mass modes $A_{m}$ are in the TeV range whereas m is of order of $M_{pl}$.

The solution for $\beta _{n}(z)$ and $\gamma _{n}(w)$ has been
presented in Eqs. (\ref{bm26b}) and (\ref{bm26c}).
The solution can be seen to include exponential factors
such as $exp\left[-\frac{5}{2}k^{2}z^{2}\right]$, $exp\left[-3\ell ^{2}w^{2}\right]$
and we see that when mass parameter $B_{m}$ is of the
order of TeV, solutions have maximum value around $z=0$ as obtained earlier
in Sec. \ref{bmbulk} as well. From the solution
of $\gamma _{n}(w)$ it is evident that the solution has maximum value
around $w=0$. Hence the bulk field being a product of
these three functions $\alpha _{p}(y)$, $\beta _{q}(z)$ and $\gamma _{r}(w)$ as shown
in Eq. (\ref{bmm4}), has mass parameter in the
TeV range and has maximum value to find the modes around $(y=\pi ,z=0,w=0)$ which is the
location of the SM brane. Also the bulk field is maximum in the SM brane; i.e., the KK modes are
most likely to be found in the TeV region as the  $A_{m}$, $B_{m}$ and $C_{m}$ are in the TeV range.
Along with the above line of arguments we could in principle
have plotted all the functions $\alpha _{p}(y)$, $\beta _{q}(z)$
and $\gamma _{r}(w)$ and for all of them we have the functions
to take maximum value at $y=\pi$, $z=0$ and $w=0$, precisely at the location of
the TeV brane.

For completeness we present the self-interactions of
the bulk scalar field in this seven-dimensional spacetime. From the
four-dimensional point of view these self-interactions
induce couplings between the KK modes. In this case also the
effective self-couplings are suppressed by the warp
factor and if the Planck scale sets the six-dimensional couplings and the low-lying
KK modes  have TeV range self-interactions. We present
the interaction term in the action with coupling parameter $\lambda$ such that
\begin{equation}\label{bmm5}
S_{int}=\int d^{4}x \int _{-\pi}^{\pi}dy \int _{-\pi}^{\pi}dz \int _{-\pi}^{\pi}dw \sqrt{G} \frac{\lambda}{M^{5m-7}}\Phi ^{2m},
\end{equation}
where the coupling $\lambda$ is of the order of unity. Then we could expand in modes and hence the
self-interactions of light KK states are given by
\begin{equation}\label{bmm6}
S_{int}=\int d^{4}x \int _{-\pi}^{\pi}dy \int _{-\pi}^{\pi}dz \int _{-\pi}^{\pi}dw R_{y}r_{z}\Re _{w}
e^{-4\sigma} \frac{\cosh^{5}(kz)}{\cosh^{5}(k\pi)} \frac{\cosh^{6}(\ell w)}{\cosh^{6}(\ell \pi)}
\frac{\lambda}{M^{5m-7}}\phi _{pqr} ^{2m}\left(\frac{\alpha _{p}}{\sqrt{R_{y}}}\frac{\beta _{q}}{\sqrt{r_{z}}}
\frac{\gamma _{r}}{\sqrt{\Re _{w}}}\right)^{2m}
\end{equation}
The effective four-dimensional coupling constants therefore are being given by
\begin{eqnarray}\label{bmm7}
\lambda _{eff}&=&\frac{8\lambda}{(MR_{y})^{m-1}(Mr_{z})^{m-1} (M\Re _{w})^{m-1}M^{2m-4}}\int _{0}^{\pi}dy
e^{-4\sigma}\alpha _{p}^{2m}
\nonumber
\\
&\times&\int _{0}^{\pi}dz \frac{\cosh ^{5}(kz)}{\cosh ^{5}(k\pi)}\beta _{q}^{2m}
\int _{0}^{\pi}dw \frac{\cosh ^{5}(\ell z)}{\cosh ^{5}(\ell \pi)}\gamma _{r}^{2m},
\end{eqnarray}
which in the large $kR_{y}$, $kr_{z}$ and $\ell \Re _{w}$ limit reduces to
\begin{eqnarray}\label{bmm8}
\lambda _{eff}\simeq &8&\lambda \left(\frac{c}{MR_{y}}\right)^{m-1}
\left(\frac{1}{Mr_{z}}\right)^{m-1} \left(\frac{1}{M\Re _{w}} \right)^{m-1}
\left(Me^{-c\pi}\frac{1}{\cosh^{2}k\pi}\frac{1}{\cosh(\ell \pi)}\right)^{4-2m}
\nonumber
\\
&\times&\int _{0}^{1}r^{4m-5}dr \left[\frac{J_{2}\left(\frac{A_{p}e^{\sigma}}{k}r\right)}{A_{p}}\right]^{2m}
\int _{0}^{\pi} dz(\beta _{q})^{2m} \int _{0}^{\pi} dw(\gamma _{r})^{2m}.
\end{eqnarray}
Hence we observe that the relevant scale for four-dimensional
physics is not the scale set by Planck scale  but
$Me^{-c\pi}\frac{1}{\cosh^{2}k\pi}\frac{1}{\cosh(\ell \pi)}$. Hence
the KK reduction lead to the TeV scale couplings by the warp
factor on the SM brane located at $(y=\pi ,z=0,w=0)$.

Now this result can easily be extended to any higher dimension
spacetime. For n extra dimensions we can write the action for the
bulk field as,
\begin{equation}\label{bmm9}
S=\frac{1}{2}\int d^{4}x\int dy\int dz\int dw \cdots \sqrt{-G}\left[G_{AB}\partial ^{A}\Phi \partial ^{B}\Phi +m^{2}\Phi ^{2}\right]
\end{equation}
where $G_{AB}$ with $A,B=\mu ,y,z,w,\cdots$ is given
by a generalization of Eq. ($\ref{bm20}$), and m is of order of $M_{pl}$.
Then the KK splitting for the bulk field can be expressed as  the following decomposition,
\begin{equation}\label{bmm10}
\Phi =\sum _{pqr\cdots}\phi _{pqr\cdots}(x)\frac{\alpha _{p}}{\sqrt{R_{y}}}
\frac{\beta _{q}}{\sqrt{r_{z}}}\frac{\gamma _{r}}{\sqrt{\Re _{w}}}\cdots
\end{equation}
Thus among these n extra dimensions, one will have
the solution given by Eq. (\ref{bm26a}), and then the other $(n-1)$ solutions
are being given by generalization of Eq. (\ref{bm26b})
such that the numerical values will be different but form of the solution
remains unaltered. For nth extra dimension $(n>1)$ the
solution for the mode can therefore be expressed as
\begin{eqnarray}\label{bmm10a}
\chi _{r}(w)&=&exp\left[-\frac{3}{2}k^{2}w^{2}\right]
exp\left[-\frac{1}{2}n k^{2}w^{2}\right] H_{\frac{\sqrt{3+n}}{2}kw}
\left(\frac{-6k^{2}-2nk^{2}+M_{r}^{2}r^{2}(1+\cosh(2k \pi))}{2k^{2}(3+n)}\right)
\nonumber
\\
&+& F_{r}exp\left[-\frac{3}{2}k^{2}w^{2}\right]\exp
\left[-\frac{1}{2}n k^{2}w^{2}\right] \nonumber
\\
&\times&
_{1}F_{1}
\left(-\frac{-6k^{2}-2nk^{2}+M_{r}^{2}r^{2}(1+\cosh(2k \pi))}{4k^{2}(3+n)},\frac{1}{2},\frac{1}{2}(3+n)k^{2}w^{2}\right)
\end{eqnarray}
Hence the bulk field as viewed by a four-dimensional observer
leads to a mass matrix whose components can be obtained by solving
the eigenvalue problem as presented by each separable
functions in the expansion given by Eq. (\ref{bmm6}). Also all
these eigenvalues have TeV scale masses and the bulk field
also has maximum value at $(y=\pi ,z=0,w=0,\cdots)$,
which is the SM brane. Hence
the standard Model particles can be taken as low-lying
Kaluza-Klein modes of a bulk field propagating in any number of extra-dimensional
spacetime.

The effective self-coupling is this case turns out to be
$$\lambda _{eff}\simeq 2^{n}\lambda \left(\frac{c}{MR_{y}}\right)^{m-1}
\left(\frac{1}{Mr_{z}}\right)^{m-1} \left(\frac{1}{M\Re _{w}} \right)^{m-1}\cdots
\left(Me^{-c\pi}\frac{1}{\cosh^{n-1}k\pi}\frac{1}{\cosh^{n-2}(\ell \pi)}\cdots\right)^{4-2m}$$

\begin{equation}\label{bmm11}
\int _{0}^{1}r^{4m-5}dr \left[\frac{J_{2}\left(\frac{A_{p}e^{\sigma}}{k}r\right)}{A_{p}}\right]^{2m}
\int _{0}^{\pi} dz(\beta _{q})^{2m} \int _{0}^{\pi} dw(\gamma _{r})^{2m}\cdots.
\end{equation}
Thus finally we have obtained the KK mass modes and their self-interactions for n extra dimensions.
We have also observed that in all these cases the KK mass modes and self-interactions are
suppressed by the warp factor near the SM brane and hence all are in TeV scale. Hence this
properties can be used to search for the TeV range KK mass modes and self-interactions in next generation colliders.
\section{Discussion}\label{bmdis}

In this paper we generalize the work presented in Ref.
\cite{Wise2} on the bulk scalar field to determine its KK modes and
the effective self-interaction in a multiple warped spacetime.
For arbitrary number of extra dimensions, we have derived the
expressions for the KK mode  masses  and their self-interactions.
Various components determining the masses are in the TeV range
because of the warp factor suppression. Moreover the bulk scalar
field has been shown to have maximum value at the SM brane. Hence
the low-lying KK modes for the bulk scalar fields lie in the TeV
range with inverse TeV self-coupling. Thus the appearance of KK
mode masses  and couplings at TeV scale are generic features of
warped dimensional models with any number of extra warped
dimensions as long as we want to resolve the gauge hierarchy
problem without introducing any hierarchical moduli. We have also
introduced the moduli stabilization mechanism in these multiply
warped models and have obtained the stabilized values for the
moduli. Then we have presented a compact and generic formula to
determine all the mass modes and their couplings for models with
any arbitrary number of warped extra dimensions. This work now can
be extended to other forms of bulk fields, which in turn may lead to
the possibility of identifying various Standard Model particles
as the low-lying KK excitation of various bulk fields, where the
small warping in multiple directions can explain mass splitting in
standard model particles as discussed in \cite{srikanth} and
\cite{Sen}. The close spacing of the low-lying KK modes along with
enhanced coupling makes it likely for them to be seen as a
series of close-lying resonances. In order to investigate the
role of KK mass modes through collider-based experiments, we
consider the interaction of various modes with thermselves, i.e.,
self-interactions, and we have obtained that all of them are
suppressed to the TeV scale by the warp factor. Also from the
numerical values of masses for low-lying KK modes, we readily
observe that the masses in the standard RS model get split into
infinite number of mass modes, with very close spacings, which is
a very interesting feature of these multiply warped models and can
be probed in future runs of LHC. The other things to be noted are
that with stronger coupling of the KK modes of the bulk scalar
field, one expects the decay widths to be larger, and thus the
peaks to be broader as we go to higher and higher dimensions. The
nature of the line shapes, therefore, will be an interesting
benchmark to distinguish between higher-dimensional and lower-dimensional 
KK signals if such excitations appear during the high-luminosity runs of the LHC.

\section*{Acknowledgements}

S.C. is funded by a SPM fellowship from CSIR, Government of India.


\end{document}